\documentclass[12pt]{iopart}
\usepackage{amssymb}
\usepackage{txfonts}
\usepackage{graphicx}

\begin{document}

\title{Angular dependence of resistivity in the superconducting state of NdFeAsO$_{0.82}$F$_{0.18}$ single crystals}

\author{Ying Jia, Peng Cheng, Lei Fang, Huan Yang, Cong Ren, Lei Shan, Chang-Zhi Gu and Hai-Hu Wen$^1$}

\address{Institute of Physics, Chinese Academy of Sciences,
Beijing 100190, China} \ead{hhwen@aphy.iphy.ac.cn}

\begin{abstract}
We report the results of angle dependent resistivity of
NdFeAsO$_{0.82}$F$_{0.18}$ single crystals in the superconducting
state. By doing the scaling of resistivity within the frame of the
anisotropic Ginzburg-Landau theory, it is found that the angle
dependent resistivity measured under different magnetic fields at a
certain temperature can be collapsed onto one curve. As a scaling
parameter, the anisotropy $\Gamma$ can be determined for different
temperatures. It is found that $\Gamma(T)$ increases slowly with
decreasing temperature, varying from $\Gamma \simeq$ 5.48 at $T=50$
K to $\Gamma \simeq$ 6.24 at $T=44$ K. This temperature dependence
can be understood within the picture of multi-band
superconductivity.

\end{abstract}

\maketitle

\section{Introduction}
Since the discovery of superconductivity at 26 K in
LaFeAsO$_{1-x}$F$_x$ \cite{sample1}, great interests have been
stimulated in the community of superconductivity. The
superconducting transition temperature was quickly raised to about
$T_c=55$ K in SmFeAsO$_{0.9}$F$_{0.1}$\cite{55KRen}, and in other
fluorine doped or oxygen deficiency samples. Moreover, hole-doped
superconductors were also successfully synthesized in
La$_{1-x}$Sr$_x$FeAsO\cite{ourhole} and Ba$_{1-x}$K$_x$(FeAs)$_2$
systems\cite{Rotter}. Many experiments have revealed that the
iron-based superconductors belong to a family with unconventional
pairing mechanism. These include the point contact tunneling
spectropy\cite{qian,shanlei}, NMR\cite{zhengguoqing}, specific
heat\cite{mugang}, lower critical field\cite{rencong} \textit{etc.}.
Meanwhile, so far many theoretical models have been proposed in
order to have a basic understanding to the mechanism of
superconductivity. As one of the basic parameters, the anisotropy
$\Gamma$ is crucial for both understanding the superconducting
mechanism and its application, because it reflects directly the
coupling strength between the "charge reservoir" LnO (Ln = La, Ce,
Pr, Nd, Gd \textit{etc}.) layers and the conducting FeAs layers.

An estimation of $\Gamma \geq$ 30 was made on
(Nd,Sm)O$_{0.82}$F$_{0.18}$FeAs polycrystals from the $c-$axis
infrared plasma frequency\cite{11}. Since the successful growth of
(Nd,Sm)FeAsO$_{0.82}$F$_{0.18}$ single crystals\cite{Smsample,fang},
precise determination of $\Gamma$ was possible. The recent study on
SmFeAsO$_{0.8}$F$_{0.2}$ single crystals revealed that the magnetic
anisotropy was temperature dependent, ranging from $\Gamma \sim$ 8
at $T\simeq T_c$ to $\Gamma \sim$ 23 at $T\simeq 0.4
T_c$\cite{singlecrystal}. Moreover, in our previous work on the
NdFeAsO$_{0.82}$F$_{0.18}$ single crystals\cite{my,chengnormal},
$\Gamma$ was calculated from the upper critical fields parallel and
perpendicular to the $ab$-plane, and was found below 5 near $T_c$.
As a further study on the superconductivity of NdFeAsO$_{1-x}$F$_x$
single crystals, here we report the angular dependence of
resistivity in the superconducting state of
NdFeAsO$_{0.82}$F$_{0.18}$ single crystals. By doing the scaling of
the resistivity based on the anisotropic Ginzburg-Landau theory, the
angle dependent resistivity measured under different magnetic fields
at a certain temperature collapse onto one curve. Thus the
anisotropy was determined for different temperatures.

\section{Experiment}

The single crystals of NdFeAsO$_{0.82}$F$_{0.18}$ were grown by flux
method at ambient pressure, which have been described in
Ref.\cite{my,fang}. The typical lateral sizes of the crystals are
about 20 $\mu m$ to 70 $ \mu m$, while thickness is about $1\sim 5$
micrometers. Electrical contacts were made using the Pt deposition
technology of a Focused-Ion-Beam (FIB) system. The upper inset of
Fig. 1 presents the image of the single crystal with Pt leads.

\begin{figure}
\includegraphics{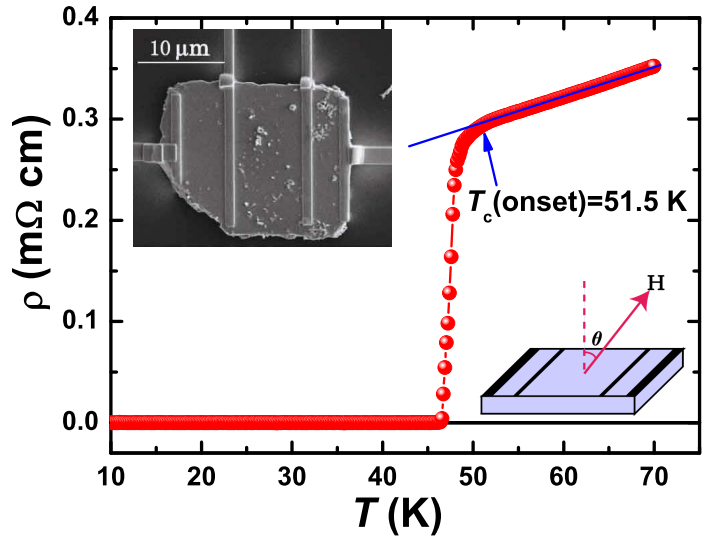}
\caption{(Color online)The resistive superconducting transition at
zero magnetic field with $T_c(onset)\simeq 51.5$K and $\triangle
T\simeq 4$K. The upper inset shows the Scanning Electron Microscope
image of the NdFeAsO$_{0.82}$F$_{0.18}$ single crystal with Pt
leads. The lower inset illustrates schematically the definition of
angle $\theta$.} \label{fig1}
\end{figure}

The angle-resolved resistivity measurements were carried out on a
Physical Property Measurement System (PPMS, Quantum Design) with
magnetic fields up to 9 T. The angle $\theta$ was varied from
$-10^{\circ}$ to $190^{\circ}$, where $\theta = 0^{\circ}$
corresponded to the configuration of $H \parallel c-$ axis and
$\theta = 90^{\circ}$ to $H \parallel ab-$ plane, respectively (as
shown in the lower inset of Fig. 1). The veracity of the angle was
ensured by the good $c$-axis orientation of the crystals, which have
been demonstrated by the X-ray diffraction (XRD)
analysis\cite{fang}. In our measurement, the current density was 150
$A/cm^2$ and the current was always perpendicular to the magnetic
field. The sample exhibited a sharp resistive superconducting
transition at $T_c(onset)\simeq 51.5$ K with $\triangle T_c\simeq 4$
K, the residual resistivity $\rho(0K)=0.13 m\Omega cm$ and
$RRR=\rho(400K)/\rho(0K)=$6.5 (shown in Fig. 1), demonstrating the
good quality of the single crystal.

\section{Results and Discussion}
The insets in Fig. 2 and Fig. 3 present four sets of data for
angular dependence of resistivity at temperatures of 44 K, 46 K, 48
K and 50 K. Taking the $\rho(\theta)$ data at $T=44$ K (the inset of
Fig. 2(a)) as an example, all the curves show a dip-like structure
with the minimum at $\theta = 90^{\circ}$ and maximum at $\theta =
0^{\circ}$ and $180^{\circ}$. The angular dependence of resistivity
is not very steep, this suggests the moderate anisotropy of the
NdFeAsO$_{0.82}$F$_{0.18}$ single crystal.

According to the anisotropic Ginzburg-Landau theory, the angular
dependence of the upper critical fields is given by
\begin{equation}
H_{c2}^{GL}(\theta)=H_{c2}^{ab}/\sqrt{\sin^2(\theta)+\Gamma^2\cos^2(\theta)},
\end{equation}
with
\begin{equation}
\Gamma = H_{c2}^{ab}(\theta = 90^{\circ})/H_{c2}^{c}(\theta =
0^{\circ})= (m_{c}/m_{ab})^{1/2} = \xi_{ab}/\xi_{c},
\end{equation}
where $H_{c2}^{ab}$ and $H_{c2}^{c}$ are the upper critical fields
with $H \parallel ab-$plane and $H \parallel c-$axis, $m_{c}$ and
$m_{ab}$ are the effective masses when the electrons move along
$c-$axis and $ab-$plane respectively, $\xi_c$ and $\xi_{ab}$ are the
coherence lengths along $c-$axis and $ab-$plane, respectively. On
this basis, Blatter \textit{et al.} developed a general scaling
approach for the angular dependence of resistivity\cite{blatter}.
Based on this method, the resistivity measured under different
magnetic fields at a certain temperature should collapse on one
curve if $x-$coordinate is properly scaled. The rescaled function
for $x-$coordinate is:
\begin{equation}
\tilde{H}=H\sqrt{\sin^2(\theta)+\Gamma^2\cos^2(\theta)},
\end{equation}

\begin{figure}
\includegraphics{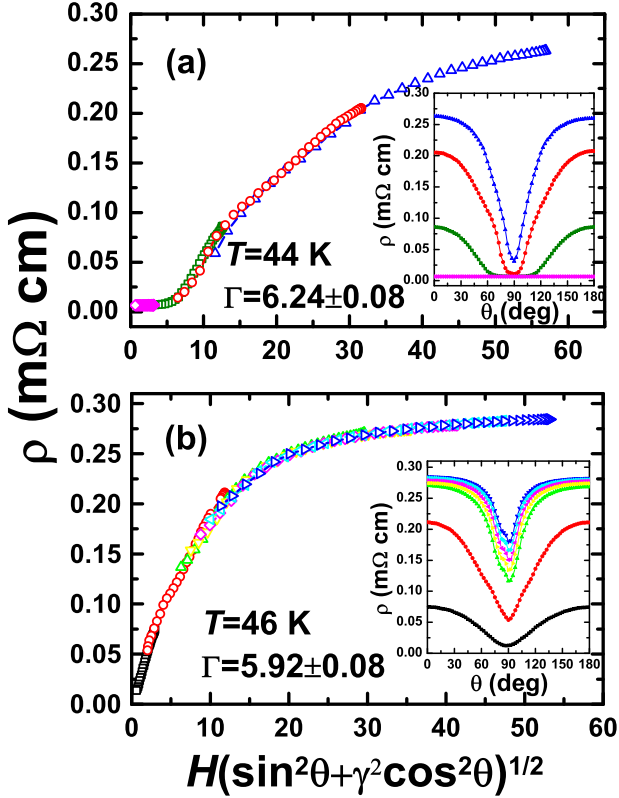}
\caption{(Color online)Main panels: the resistivity as a function of
$\tilde{H}$ (see text) for (a) $T=44$ K and (b) $T=46$ K. Insets:
the angular dependence of magnetoresistivity at (a) $\mu_{0}H=$ 0.5,
2, 5, 9 T and (b) $\mu_{0}H=$ 0.5, 2, 5, 6, 7, 8, 9 T. The
anisotropy parameters are found to be 6.24 $\pm$ 0.09 for $T=44$ K
and 5.92 $\pm$ 0.09 for $T=46$ K} \label{fig1}
\end{figure}

\begin{figure}
\includegraphics{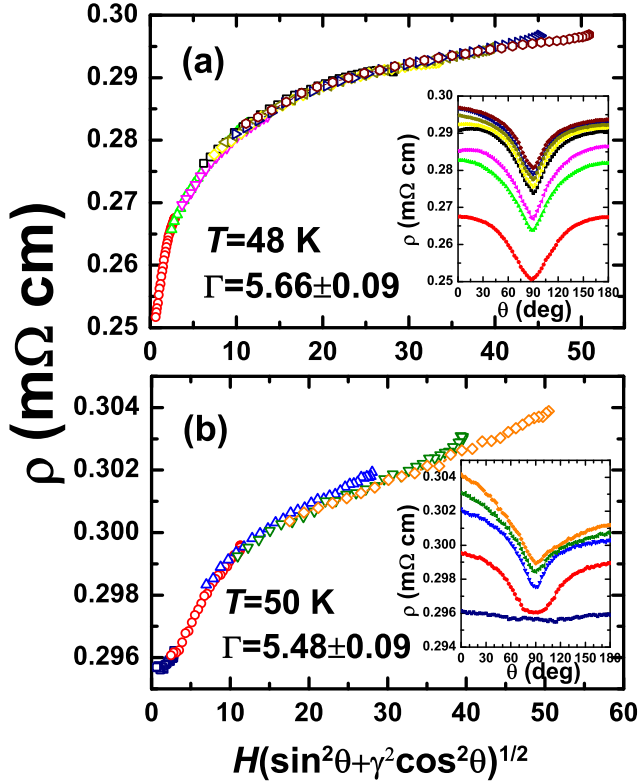}
\caption{((Color online)The resistivity versus $\tilde{H}$ (see
text) for (a) $T=48$ K and (b) $T=50$ K. The insets present the
angular dependence of magnetoresistivity at (a) $\mu_{0}H=$ 0.5, 2,
3, 5, 6, 7, 8, 9 T and (b) $\mu_{0}H=$ 0.5, 2, 5, 7, 9 T. The
scaling parameters  are $\Gamma= 5.48 \pm$ 0.08 for $T=48$ K and
$\Gamma= 5.66 \pm$ 0.08 for $T=50$ K}\label{fig1}
\end{figure}

By adjusting $\Gamma$, excellent scaling curves are obtained for
each set of data at different temperatures, shown in the main panels
of Fig. 2 and Fig. 3. In this treatment, only one variable $\Gamma$
is employed as the fitting parameter, so the value of $\Gamma$ is
more reliable comparing with that determined from the ratio of
$H_{c2}^{ab}$ and $H_{c2}^{c}$, which might be affected by the
criterion of $H_{c2}$ determination. The anisotropy parameter
$\Gamma$ takes 6.24$\pm$0.08 for $T = 44$ K, 5.92$\pm$0.08 for $T =
46$ K, 5.66$\pm$0.09 for $T = 48$ K and 5.48$\pm$0.09 for $T = 50$
K. Such values are consistent with that we estimated from
$H_{c2}^{ab}/H_{c2}^{c}$\cite{my}, but smaller than the values
obtained on SmFeAsO$_{0.8}$F$_{0.2}$ single crystals using a torque
technique\cite{singlecrystal}. It is found that, as plotted in Fig.
4, the anisotropy increases slowly with decreasing temperature. The
similar behavior of $\Gamma(T)$ was also observed in
SmFeAsO$_{0.8}$F$_{0.2}$ single crystals\cite{singlecrystal}, which
was attributed to the two-gap scenario\cite{twogap,multiband}. In
addition, it should be noted that the good scaling behavior suggests
a field-independent anisotropy in our measuring range.

\begin{figure}
\includegraphics{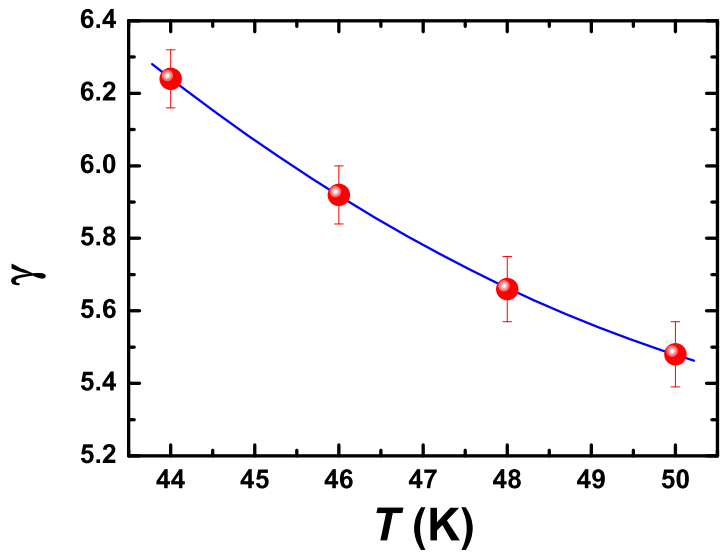}
\caption{(Color online)Temperature dependence of anisotropy
parameter $\Gamma$. The line is guided to eyes.} \label{fig4}
\end{figure}

For comparison, in MgB$_2$, a typical two-band superconductor, clear
deviations have been observed between the experimental
$H_{c2}(\theta)$ and the anisotropy G-L description. Such a
phenomenon was attributed to the different anisotropy factors of the
two bands\cite{MgB2twoband}. Whereas, no such deviation is found in
our experiment, so we suggest that single band or multi-band with
similar anisotropy parameters should exist in the
NdFeAsO$_{0.82}$F$_{0.18}$ superconductors.

\section{Summary}
In conclusion, we have investigated the angle dependent resistivity
in the superconducting state of NdFeAsO$_{0.82}$F$_{0.18}$ single
crystals. It is found that the $\rho(\theta,H)$ data measured at a
certain temperature can be described by a scaling law based on the
anisotropic Ginzburg-Landau theory. Thus the values of anisotropy
are extracted for different temperatures. It is found that
$\Gamma(T)$ increases with decreasing temperature, which was
attributed to the multi-band effect.

\section{Acknowledgements}
This work is supported by the Natural Science Foundation of China,
the Ministry of Science and Technology of China (973 project No:
2006CB60100, 2006CB921107, 2006CB921802), and Chinese Academy of
Sciences (Project ITSNEM).

\end{document}